\documentstyle[aps,twocolumn,floats,prl,psfig]{revtex}

\newcommand{\ApJL}{Astrophys. J. Lett.}
\newcommand{\ApJ}{Astrophys. J.}

\newcommand{\PRD}{Phys. Rev. D}
\newcommand{\MNRAS}{Mon. Not. R. Astron. Soc.}
\newcommand{\PASA}{Publ. Astron. Soc. Aust.}

\newcommand{\bx}{\hat{\bf x}}
\newcommand{\br}{\hat{\bf R}}
\newcommand{\bl}{\hat{\bf L}}
\newcommand{\by}{\hat{\bf y}}

\begin{document}
\twocolumn[\hsize\textwidth\columnwidth\hsize\csname
@twocolumnfalse\endcsname

\title{Is the Cosmic Microwave Background Circularly Polarized?}
\author{Asantha Cooray$^1$, Alessandro Melchiorri$^2$ and Joseph Silk$^2$}
\address{$^1$Theoretical Astrophysics, California Institute of Technology, Pasadena, CA 91125. E-mail: asante@caltech.edu\\
$^2$Astrophysics, Denys Wilkinson Building, University of Oxford, Oxford, OX 3RH. E-mail:(melch,silk)@astro.ox.ac.uk}


\maketitle

\begin{abstract}
The primordial anisotropies of the cosmic microwave background (CMB) are 
linearly polarized via Compton-scattering. The Faraday conversion process during the propagation of 
polarized CMB photons through regions of the large-scale structure
containing magnetized relativistic plasma, such as galaxy clusters, will lead to a circularly polarized contribution.  
Though the resulting Stokes-V parameter is of order 10$^{-9}$ at 
frequencies of $10$ GHz, the contribution can potentially reach the level of total Stokes-U at low frequencies due to the 
cubic dependence on the wavelength. In future, the detection of circular polarization of CMB can be used as a potential
probe of the physical properties associated with relativistic particle populations in  large-scale structures.
\end{abstract}
\vskip 0.5truecm


]

The CMB anisotropies are expected to be linearly polarized
by anisotropic Compton scattering around the epoch of recombination
\cite{kaiser}. This linear polarization field has been
widely discussed in the literature \cite{HuWhi97} and its accurate measurement
can ultimately shed new light on the thermal history of the universe and on the primordial gravitational-wave
background. It is also well established that, in the standard scenario,
no relevant circular polarization should be present.
For these reasons, many of the present and near future CMB
experiments like MAP and Planck Surveyor have not been designed for a detection of circular polarization.

It is therefore quite probable that in the near future,
in spite of a continuous incremental knowledge on linear
CMB polarization, the experimental bounds on the circular component
will not drastically improve from those of
early observations \cite{lubin}. Since post-Planck CMB polarization experiments are already
under study \cite{peterson}, it is extremely timely to address the question whether
CMB is  circularly polarized and what physical information can be extracted from its measurement.
This {\it letter} represents a discussion in this direction.

In order to understand the CMB polarization field,
we make use of the Stokes parameters \cite{Jac75}. In the case of a propagating wave in the $z$ direction, 
${\bf  E} = (E_x e^{i\phi_x} \bx + E_y e^{i\phi_y} \by) e^{-i\omega t}$, 
with amplitudes $E_x$ and $E_y$ in the $x$ and $y$-directions with phases
$\phi_x$ and $\phi_y$, respectively, we can write the Stokes parameters 
as time averaged quantities: 
\begin{eqnarray}
I &\equiv& \langle E_x^2 \rangle + \langle E_y^2 \rangle \, , \nonumber \\
Q &\equiv& \langle E_x^2 \rangle - \langle E_y^2 \rangle \, , \nonumber \\
U &\equiv& \langle 2E_xE_y \cos (\phi_x - \phi_y )\rangle \, , \nonumber \\
V &\equiv& \langle 2E_xE_y \sin (\phi_x - \phi_y )\rangle \, .
\label{eqn:stokes}
\end{eqnarray}
Note that the total intensity of the radiation is given by the 
Stokes-I parameter, while for unpolarized
radiation $Q=U=V=0$. The linearly polarized radiation is defined by  non-zero values for $Q$ and/or $U$.
These latter two Stokes parameters form a spin-2 basis; A rotation of the coordinate system, by
an angle $\theta$, leads to a new set of parameters for the same radiation field
given by $(\bar Q \pm i \bar U) =  (Q \pm iU) e^{2 i \theta}$.  This coordinate dependence is avoided in the
literature by introducing a new set of orthonormal basis with a part containing the gradient of a scalar
field, called grad- or E- modes, and a part containing the curl of a vector field, called curl- or B-modes 
\cite{KamKosSte97a}. Note that the Stokes-V parameter,  
which is coordinate-independent similar to Stokes-I,  defines the extent to which radiation is circularly polarized.

Though there may not be a physical mechanism to generate a Stokes-V contribution at the
last scattering surface, the radiation detected today, however, is not exactly the field
that last scattered. During the propagation from the last scattering surface to us, 
CMB photons encounter large-scale structures and undergo significant 
changes due to effects related to  structure formation \cite{Coo02}.
These modifications include a regeneration of new anisotropies, such as through 
the Sunyave-Zel'dovich (SZ; \cite{SunZel80}) effect involving
the inverse-Compton scattering of CMB photons via hot electrons in galaxy clusters.
The transit of CMB photons also leads to modifications to the polarization signal. 
For example, the gravitational lensing deflection of
the  CMB  propagation directions leads to a transfer of power from the dominant E-mode of  
polarization to the B-mode \cite{ZalSel98}. 

In the case of CMB, galaxy clusters present potential sources where interesting polarization modifications occur.
In addition to the presence of thermal electrons, large-scale diffuse synchrotron emission towards galaxy 
clusters suggests the presence of magnetic fields \cite{Claetal01}. The propagation of radiation through such magnetized plasma
lead to the well-known modification involving the rotation of linear polarization between Stokes-Q and -U parameters 
via the Faraday rotation (FR) \cite{RybLig79}. The FR comes about from the fact that normal modes of propagation in a magnetized 
plasma are circularly polarized. Instead of the description given in 
equation~(\ref{eqn:stokes}), then, it is useful to consider the
linearly polarized radiation field, in terms of a superposition of equal left and right-hand circularly polarized contributions. 
We can write ${\bf  E} = (E_R e^{i\phi_R} \br + E_L e^{i\phi_L} \bl) e^{-i\omega t}$ with unit vectors 
for the right-, $(\br)$, and left-, $(\bl)$, hand polarized waves as $(\bx \mp i\by)/\sqrt{2}$, respectively. 
In terms of this redefinition of the radiation field, the Stokes-Q and -U parameters are
\begin{eqnarray}
Q &\equiv& \langle 2E_RE_L \cos (\phi_R - \phi_L )\rangle \, , \nonumber \\
U &\equiv& \langle 2E_RE_L \sin (\phi_R - \phi_L )\rangle \, .
\label{eqn:circstokes}
\end{eqnarray}
The right- and left-hand circularly polarized waves travel through the magnetized medium 
with different phase velocities introducing an additional phase shift involving 
$(\phi_R-\phi_L)$. This phase shift mixes Stokes-Q and -U parameters such that
\begin{eqnarray}
\dot{Q} = -2 U \frac{d \Delta \phi_{\rm FR}}{dt}   \quad \quad  {\rm and} \quad \quad
\dot{U} = 2 Q \frac{d \Delta \phi_{\rm FR}}{dt} \, ,
\end{eqnarray}
where the overdot is the derivative with respect to time, $t$.

The associated mixing is described via the rotation measure angle
\begin{eqnarray}
\Delta \phi_{\rm FR} &=& \frac{e^3\lambda^2}{2 \pi m_e^2c^4} \int dl n_e(l) {\bf B} \mu \, , \nonumber \\
&\approx& 8 \times 10^{-2}\; {\rm rad}\; (1+z)^{-2} \left(\frac{\lambda_0}{1\; {\rm cm}}\right)^2 \nonumber \\
&& \quad \times \int \frac{dl}{1\; {\rm kpc}} \left(\frac{n_e}{0.1\; {\rm cm}^{-3}}\right)  \left(\frac{{\bf B}}{10\; \mu {\rm G}}\right) \mu  \, .
\label{eqn:fr}
\end{eqnarray}
Here, $\mu$ is the cosine of the angle between the line of sight direction and 
the magnetic field, ${\bf B}$, in galaxy clusters, $n_e$ is the
number density of electrons, $\lambda=\lambda_0(1+z)^{-1}$ is the wavelength of radiation
with wavelength today given by $\lambda_0$, while rest of the parameters 
have generally known values.

The FR effect on CMB anisotropies has been considered in the literature as a possible way to
generate a Stokes-U contribution from the dominant Stokes-Q contribution 
both for a primordial \cite{LoeKos96} and galaxy cluster 
\cite{Taketal01} magnetic fields. In the case of galaxy cluster magnetic fields, 
the observed rotation measures up to 250 radians m$^{-2}$ in nearby massive clusters 
\cite{Claetal01}, suggest that the CMB linear polarization is rotated, on average, by 
an angle of order $10^{-1}$ radians at an observed frequency of few GHz.
Thus, FR can potentially generate a Stokes-U contribution which is of order $10^{-7}$ in fractional temperature, 
$\Delta T_{\rm pol}/T_{\rm CMB}$ \cite{FT}, 
from the dominant Stokes-Q contribution with an rms of order $10^{-6}$ at 10 GHz \cite{Taketal01}. 

The propagation of radiation has already been discussed in the literature as an 
explanation for the circular polarization observed towards certain extragalactic radio sources \cite{Pac73}.
A similar situation applies to the CMB. It is now well known that galaxy clusters also contain populations of 
relativistic particles; The observed hard X-ray and the extreme ultraviolet emission require their presence \cite{Sar00}. 
In a magnetized plasma containing highly relativistic electrons, the normal modes of propagation, however, are not
perfectly circular but rather linear.  The linear modes of propagation are also encountered when
one is dealing with  uniaxial crystals and so-called quarter wave plates which are configured to 
convert fully linear polarized radiation to a circularly polarized
contribution.  In the case of relativistic plasmas embedded in a magnetic field, the conversion of linear polarization to
circular polarization can be described under the
formalism of  generalized Faraday rotation \cite{Pac73}.

The effect is essentially same as the well known FR effect, except the propagation is considered
in terms of the linear modes rather than the circular modes. Even if the plasma is not relativistic, 
under generalized Faraday rotation, the modes of propagation are linear when the radiation 
field is propagating more or less close to a perpendicular direction to the magnetic field. 
The conversion can  be understood in terms of the description of the radiation 
field given in equation~(\ref{eqn:stokes}) for linear waves instead of the decomposition to
circular states in equation~(\ref{eqn:circstokes}).
The difference in phase velocities now lead to a mixing between the Stokes-U and -V parameters with the
addition of a phase shift to $(\delta_x-\delta_y)$. This mixing is similar to the ones involving
Stokes-Q and -U under the normal FR with the introduction of a phase shift to $(\delta_R-\delta_L)$.

Since Stokes-V is effectively zero for the incoming radiation, 
the outgoing radiation contains a contribution to the Stokes-V and, therefore,
the effect is generally described in the literature as the Faraday conversion (FC) \cite{JonOde77}. We
can write the converted Stokes-V contribution as
\begin{equation}
\dot{V} = 2 U \frac{d \Delta \phi_{\rm FC}}{dt} \, .
\end{equation}
The rotation measure angle related to FC in a magnetized relativistic plasma, 
analogous to the rotation measure associated with the FR effect is \cite{Pac73},
\begin{eqnarray}
&&\Delta \phi_{\rm FC} = \frac{e^4\lambda^3}{\pi^2m_e^3c^5} \left(\frac{\beta-1}{\beta-2}\right)
\int dl n_r(l) \gamma_{\rm min} |{\bf B}|^2 (1-\mu^2) \, , \nonumber \\
&\approx& 3 \times 10^{-7}\; {\rm rad}\; (1+z)^{-3} \left(\frac{\lambda_0}{1\; {\rm cm}}\right)^3 \left(\frac{\beta-1}{\beta-2}\right)_{\beta=2.5} \nonumber \\
&\times& \int \frac{dl}{1\; {\rm kpc}} \left(\frac{n_r}{0.1\; {\rm cm}^{-3}}\right)  \left(\frac{\gamma_{\rm min}}{300}\right) \left(\frac{|{\bf B}|}{10\; \mu {\rm G}}\right)^2 (1-\mu^2)\, .
\label{eqn:fc}
\end{eqnarray} 
Here, $n_r$ is the number density of relativistic particles and $\beta$ defines the power-law distribution of
the particles, in terms of the Lorentz-factor $\gamma$, such that
\begin{equation}
N(\gamma) = N_0 \gamma^{-\beta} \, ,
\end{equation}
between $\gamma_{\rm min} < \gamma < \gamma_{\rm max}$. 
Other parameters are same as the ones defined in equation~(\ref{eqn:fr}). 
A comparison of equations~(\ref{eqn:fr}) and (\ref{eqn:fc})
reveals that while the FR is proportional to the square of the observed wavelength, the FC 
scales as the cube of the wavelength. Thus, one finds a stronger wavelength dependence for the
FC when compared to the rotation. 

A quick estimate for the two suggests that the FC is at least $10^{5}$ orders of magnitude smaller than the FR effect.
Unlike FR, however, the FC simply depends on the number density of relativistic particles; a consequence of this
is that an equal mixture of positive and negative particles will contribute to circular polarization conversion while
there will be no rotation associated with linear polarization. Also, conversion depends on the square of the 
amplitude of the magnetic field and not the magnetic field itself.  Thus, in certain favorable astrophysical conditions, 
the FC to circular polarization can be significant leading to a measurable contribution to the Stokes-V parameter.
As a potential source of conversion between Stokes-U to V, we will consider galaxy clusters, as there is
some evidence for populations of relativistic particles in these massive objects.

The extent to which galaxy clusters convert Stokes-Q to a Stokes-U parameter under FR
has already been discussed in the literature \cite{Taketal01}. The rotation effect 
encountered here depends on the properties of the
magnetic field and the distribution of thermal electrons, both of which are now well known for clusters through 
 X-ray and synchrotron emission observations. The radio measurements of
FR through intracluster gas indicate magnetic fields 
or order  tens of microGauss towards nearby massive galaxy clusters \cite{Claetal01}.
The extreme ultraviolet and the hard X-ray  emission observed towards certain clusters suggest 
the presence of relativistic electrons with bulk Lorentz factors of order $\sim$ 300 and $\sim 10^4$,
respectively \cite{Sar00}. The calculations that attempt to explain these observations generally
suggest relativistic populations with a spectrum $N(\gamma) \propto \gamma^{-\beta}$ 
where $\beta \sim 2.3$ and as steep as $\sim$ 3.3.

Assuming reasonable parameters for galaxy clusters with $B=10 \mu$G, a path length of 1 Mpc, 
which is a typical size for a massive cluster, $\gamma_{\rm min}=100$ for relativistic particles,
and an observed frequency of 10 GHz, we estimate $\Delta \phi_{\rm FC} \sim {\rm few} \times 10^{-3}$. 
With a typical rms contribution of order $10^{-6}$ to the incoming CMB polarization that propagate through galaxy 
clusters, the outgoing radiation should contain a circular polarization of order $10^{-9}$
at scales corresponding to galaxy clusters. Note that this estimate is highly uncertain
by at least two orders of magnitude both due to the unknown number density of relativistic particles and
the Lorentz-factor distribution of these particles. Since one expects a 
contribution to the Stokes-V parameter when the radiation is propagating nearly perpendicular to the magnetic field,
the final contribution not only depends on the magnitude of the magnetic field, 
but also on detailed physical properties such as the spatial distribution.
Due to the $\lambda^3$ dependence on the wavelength, the Stokes-V contribution can potentially
reach the maximal Stokes-U contribution at low frequencies of 1 GHz and below.

\begin{figure}[!t]
\centerline{\psfig{file=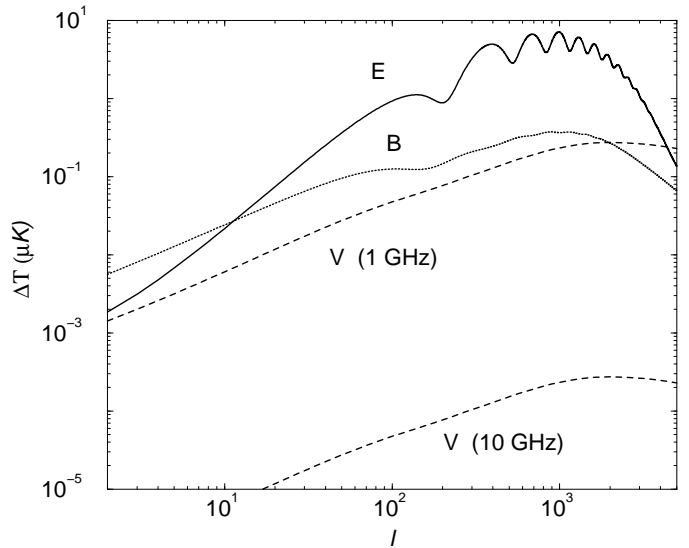,width=3.5in,angle=-90}}
\caption{
The flat-band power spectra of CMB polarization ($\Delta T =
T_{\rm CMB} \sqrt{l(l+1)/2\pi C_l})$. We show contributions to the E-mode
(solid line), B-mode (dotted line) and estimates for the Stokes-V mode at
10 (dashed line) and 1 GHz (long-dashed line). The contribution to B-modes
contains two parts involving gravitational-waves at large angular scales
and gravitational lensing effect at small angular scales. The shown V-mode
contributions should be considered reasonable given uncertainties
associated with relativistic populations. Given the significant wavelength
dependence, $\lambda^3$ in this plot, low-frequency observations are
desirable to detect the circular polarization contribution.} 
\label{fig:cl}
\end{figure}

We can extend the approach presented in Ref. \cite{Taketal01}, following the so-called halo model \cite{CooShe02}, to
calculate the expected angular power spectrum of the Stokes-V contribution. The Stokes-V correlation 
is simply the product of correlation functions involving Stokes-U contribution and the
FC rotation measure: $C_V(\theta)=C_U(\theta)C_{\rm FC}(\theta)$.
The correlation function associated with the Stokes-U contribution can be written as \cite{KamKosSte97a}: 
\begin{eqnarray}
&&C_U(\theta) =  \nonumber \\
&&\int \frac{l dl}{2\pi} \left\{\frac{C_l^{\rm EE}}{2} \left[J_0(l\theta)-J_4(l\theta)\right]+
\frac{C_l^{\rm BB}}{2} \left[J_0(l\theta)+J_4(l\theta)\right]\right\} \, .
\end{eqnarray}
We assume a total contribution to the B-mode power spectrum, $C_l^{\rm BB}$, 
from both gravitational waves, with a tensor-to-scalar ratio of 0.1,
 and gravitational lensing conversion of E to B-mode.  
Since lensing, effectively, happens at redshifts greater than 1 \cite{CooKes02}, while FC happens in
massive clusters at redshifts less than 1, it is unlikely that we have overestimated the total linear polarization 
contribution that can be converted to the circular polarization. 
To calculate $C_{\rm FC}(\theta)$, we use a halo distribution with masses
greater than $10^{14}$ M$_{\odot}$ with the assumption that the distribution of relativistic particles in these
halos trace the gas distribution and the magnetic field in each cluster is constant, which in this case we set at
10 $\mu$Gauss. We summarize our results in figure~\ref{fig:cl}.

In addition to galaxy clusters as discussed above,  large-scale shocks involved with the formation of
structures, including galaxy clusters, could be significant sources of magnetized plasma in which FC 
may be efficient.  Due to the strong dependence on wavelength, the Faraday conversion effect can
easily be identified and separated from other contaminant contributions such as radio 
point sources that may dominant the polarization signal at low frequencies. An additional, and 
possibly important, source of circular polarization is the conversion associated with a primordial magnetic field. 
Though current observations limit a primordial magnetic field to be at the level of $10^{-3}$ $\mu$G today 
\cite{Val90},  the evolution of the field as $(1+z)^2$ will lead to a significant contribution during the 
recombination era.  A limit on the circular polarization at early times can be used as a way to put a 
reliable limit on the large scale primordial magnetic field. We will return to this subject in detail in the future.

Though our first estimate on the level of CMB circular polarization due to galaxy clusters 
is smaller than the contribution to the linear polarization, 
observational studies on circular polarization are clearly warranted.  
As discussed, the Faraday conversion
involves the presence of relativistic particles and their detailed physical properties.
Any detection of the circular polarization will allow a probe of these relativistic 
populations in the large scale structure and the associated magnetic fields. 
Furthermore, a detection of a $V$ contribution  comparable in signal to the $Q$ and $U$ 
modes would be nearly impossible to explain in the standard scenario, requiring 
the introduction of new physics or, again, providing an useful check for systematics.

Though the required sensitivity level to detect circular polarization is beyond what 
is allowed by current instrumental techniques, with new detector 
technologies and observational methods, it is likely that the required level 
of sensitivity will be reached in the future. 
As with Faraday rotation, the ultimate limitation will be contamination from 
foreground sources. At frequencies around $10$ GHz and below, the synchrotron emission
from galaxy is dominant with respect to the CMB. Measurements at $1$ GHz shows a circular polarization
from this foreground at level of $10^{-4}$ \cite{kellermann}.
However, the circular polarization contribution associated 
with galaxy clusters could be separated out by cross-correlating with large-scale 
structure data and/or by the different frequency dependence.

Though most CMB experiments are not sensitive to the Stokes-V contribution, interferometric arrays 
detect this contribution. Therefore, the anticipated polarization data from interferometers
such as  DASI and CBI, at 30 GHz, will provide useful upper limits on the $V$-mode contribution 
at the order of a few $\mu K$.  Given the strong wavelength dependence, however, any attempt to detect circular polarization 
should be considered at low frequencies; In this respect, the upcoming Square Kilometer Array (SKA) and the 
Low Frequency Array (LOFAR) may provide interesting results in this direction.
The relevant astrophysical uses associated with CMB circular polarization 
clearly motivate future observational programs for a positive detection.

This work was supported at Caltech by the Department of Energy and the Sherman Fairchild foundation.
AC thanks the Oxford astrophysicists for their hospitality during a brief visit where this work was initiated.
AM, supported by PPARC, thanks Francesco Melchiorri and Chris O'Dell for valuable comments.


\begin{thebibliography}{99}
\bibitem{FT} We quote the polarization in terms of fractional difference in
brightness temperature  as both the temperature and polarization anisotropies have the same thermal spectrum.
\end{thebibliography}
\end{document}